\documentstyle[nato,numreferences,epsf,amsfonts]{crckapb} 

\newcommand{\gfive}{\gamma_{5}}     
\newcommand{\psibar}{\overline{\psi}}    
\newcommand{\Abar}{\overline{A}}    
\newcommand{\cD}{{\cal D}}    
\newcommand{\gmu}{\gamma_{\mu}}

\def\identity{{\Bbb I}}

\begin{opening}
\title{SOME ANSWERED AND UNANSWERED QUESTIONS\protect\\
       ABOUT THE STRUCTURE OF THE SET OF FERMIONIC\protect\\
       ACTIONS WITH GWL SYMMETRY}

\author{IVAN HORV\'ATH}
\institute{Department of Physics\\
University of Virginia\\
Charlottesville, VA 22903\\
USA
}

\end{opening}

\begin{document}


\section{Introduction}

In this talk I will briefly discuss several issues that I thought 
about since Ginsparg-Wilson (GW) relation and  
Ginsparg-Wilson-L\"uscher (GWL) symmetry became popular topics
in lattice field theory. Most of these issues are not resolved
to my satisfaction (if at all), which actually makes them an
appropriate material to discuss at a workshop like this.

In lattice field theory we typically want to use some finite 
or countably infinite set of variables to define, as a sequence 
of approximations, the theory which formally involves a continuous
infinity of variables. The most important guide to do this, both
correctly and efficiently, are the symmetries. The dynamics of 
theories relevant in particle physics (such as QCD) is crucially 
driven by {\it (i)} Poincar\'e symmetry {\it (ii)} gauge symmetry,
and {\it (iii)} chiral symmetries. 

Obviously, the lattice counterparts of these do not involve precisely 
the same transformations, since they act on a different set of degrees
of freedom. The goal is rather to choose the discrete set of variables
and the set of symmetry conditions so that the dynamics is constrained
in a way analogous to that in the continuum. While this is quite non-unique,
we usually stick to very definite choices. Thus, trying to account for 
at least some of the Poincar\'e invariance, the variables are usually 
associated with the hypercubic lattice structure, and their
Euclidean dynamics is required to respect its symmetries. With gauge 
invariance in mind, it is most common to associate fermionic variables 
with sites and gauge group elements with links of the
lattice, and to form actions built out of closed gauge loops or open 
gauge loops with fermionic variables at the ends. While in this setup
it is trivial to restrict the actions further by requiring the invariance 
under the on-site $\gamma_5$ rotation (naive chiral symmetry), for well
known reasons, the resulting set of actions is just too small to define
the theories we want. 

This situation is shown in Fig.~1, where the set $A$ represents
acceptable fermionic actions, quadratic in fermionic variables, and with 
``easy symmetries''. The subset $A^L$ of local actions is usually 
considered (with at least exponentially decaying couplings at large 
distances in arbitrary gauge background), because of the fear of 
non-universality in the non-local case. The problems with naive chiral 
symmetry are reflected by the fact that there is no intersection of the 
subset of symmetric actions $A^C$ with the subset of doubler-free 
actions $A^{ND}$ on the local side of the diagram. This is a consequence 
of the Nielsen-Ninomiya theorem~\cite{Nie81A}.
 
A possible clean resolution of this is contained in a proposition
that lattice theory, for which the chirally nonsymmetric part of the 
propagator is local, is in virtually all important aspects as good as 
the one with chirally nonsymmetric part being zero~\cite{Gin82A,Has98B}. 
This is plausible, because the ``important aspects'' are typically 
associated with properties of fermionic correlation functions at large 
distances. These, in turn, 
depend crucially on the long distance behaviour of the propagator,
hence the significance of the above property. The actions satisfying
this requirement became known as GW actions, and if we represent the
elements of $A$ by corresponding Dirac kernels $D$, then we have
\begin{displaymath}
  A^{GW}\; \equiv\; \{\, D \in A \,:\, (D^{-1})_N \; 
  \mbox{{\tt is local}}\,\} 
  \qquad\;
  (D^{-1})_N \;\equiv\; {1\over 2} \gfive \{\gfive,D^{-1}\} 
\end{displaymath}
GW kernels are not particularly generic. For free Wilson--Dirac operator
in Fourier space we have for example 
\begin{displaymath}
  \bigl( D_W^{-1}\bigr)_N \;=\; { \sum_\mu 1-\cos p_\mu \over
   {(\sum_\mu 1-\cos p_\mu)^2 + \sum_\mu \sin^2 p_\mu}}
   \;\identity        
\end{displaymath} 
where $\identity$ is the identity matrix in spinor space. The second
partial derivatives of the scalar function in the above expression
are directional, implying that the operator is non-local. The chirally
nonsymmetric part of the propagator affects the long distance physics, 
and chiral properties of Wilson--Dirac operator are bad.

\begin{figure}
\centering
\epsfxsize=9.0cm
\epsffile{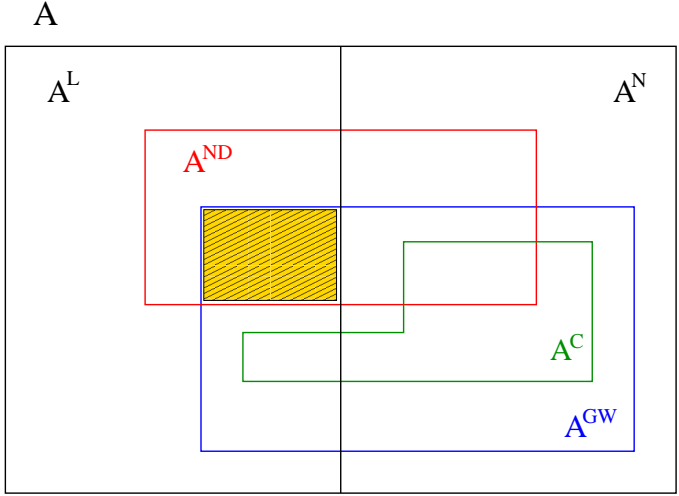}
\caption{Lattice fermionic actions. Indicated are sets with following
         properties: $A$ - hypercubic symmetries, gauge invariance,
         relativistic ``naive'' continuum limit; $A^L$ - local, 
         $A^N$ - nonlocal, $A^{ND}$ - no doublers, $A^C$ - chiral
         symmetry, $A^{GW}$ - GWL symmetry.}
\vspace{-0.0cm}
\end{figure}

The reason why the above considerations are exciting is that it appears
that fermion doubling is not a definite property of local GW 
actions~\cite{Has98C,Neu98B,Her98A}. In other words, 
$A^{GW} \cap A^{ND} \cap A^L \not= \emptyset$, 
as indicated on Fig.~1. Apart from a conjectured 
existence of this intersection, not much is known about the structure of 
the set $A^{GW}$. The relevant interesting questions include 
the following: Are there any useful definite properties of the set 
$A^{GW}$ and the set $A^{GW} \cap A^{ND}$? Can we classify all GW actions 
by some useful characteristics? How simple can GW actions be? What is a
good definition of ``simple'' for these actions? In what follows, I will
discuss certain issues that are relevant to these kind of questions.

\section{Non-Ultralocality of GWL Transformations}

There is one fully general result here that reveals the inherent property
of GW actions and the nature of symmetry they share~\cite{Hor99A}. 
Considering the GWL transformations
$ \delta \psi = i\theta\gfive (\identity- R D) \psi$, 
$\; \delta \psibar = \psibar\, i\theta (\identity-D R)\gfive$, 
such that $[R,\gfive]=0$, 
it is well known that the set $A^{GW}$ can be alternatively 
defined through the symmetry principle~\cite{Lus98A}, i.e.
\begin{displaymath}
    A^{GW}\; \equiv\; \{\, D \in A, \; \exists\; R \; \mbox{{\tt local}}
    \;:\; \delta (\psibar D \psi)=0\,\} 
\end{displaymath}
The following result can be proved~\cite{Hor99A}:
\begin{itemize}
   \item[] {\it If $D \in A^{GW}$, then the corresponding infinitesimal 
    GWL transformation couples variables at arbitrarily large lattice 
    distances, except when $R = 0$ (standard chiral symmetry)}.
\end{itemize}
This is equivalent to non-ultralocality of 
$\; \cD \;\equiv\; 2 R D = 2 (D^{-1})_N D$, 
assigned to any $D \in A^{GW}$, except when $D \in A^C$.
 Ref.~\cite{Hor99A} actually deals 
in detail with the physically relevant case of local elements of $A^{GW}$, 
but it is in fact true for all elements.

\begin{figure}
\centering
\epsfxsize=9.0cm
\epsffile{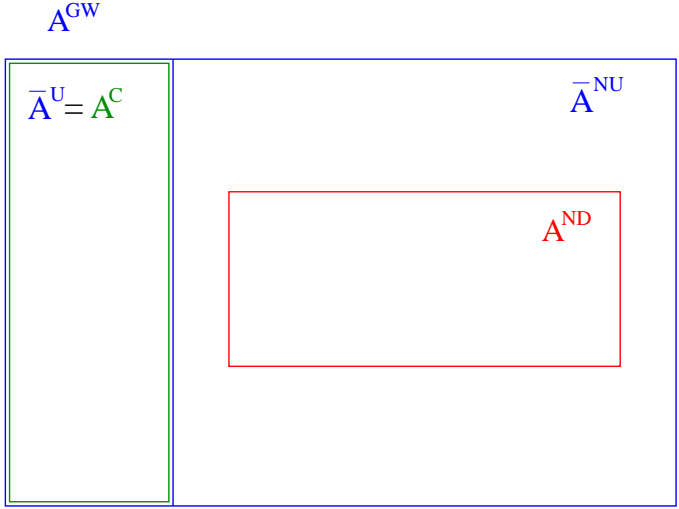}
\caption{Ultralocality properties of infinitesimal GWL symmetry 
         transformations for the set $A^{GW}$. Subset $\Abar^U$ 
         has ultralocal and $\Abar^{NU}$ non-ultralocal transformations.}
\vspace{-0.0cm}
\end{figure}

The above theorem on {\it ``weak non-ultralocality''} is represented on
Fig.~2, where the set $A^{GW}$ is split into the parts with ultralocal
GWL transformation ($\Abar^U$), and non-ultralocal GWL transformation
($\Abar^{NU}$). Since $\Abar^U = A^C$, this means that there is a sharp
discontinuity in the set $A^{GW}$. Naively, $A^C$ represents a smooth 
limit in $A^{GW}$ as the chirally nonsymmetric part of the propagator 
completely vanishes. However, while chiral transformation only mixes 
variables on a single site, the nontrivial infinitesimal GWL symmetry 
operation requires rearrangement of infinitely many degrees of freedom 
(on infinite lattice). This is a necessary requirement to achieve the 
delicate goal of preserving chiral dynamics, while keeping doublers away.     

\section{Non-Ultralocality of GW Actions}

From the theorem on weak non-ultralocality it follows that, except for the
subset $A^C$, there are no ultralocal elements of $A^{GW}$ for which 
$R = (D^{-1})_N$ is ultralocal~\cite{Hor98A,Hor99A}. Apart from conceptual 
value, this obviously has some serious practical consequences for both 
perturbation theory and numerical simulations.
While the above subset of $A^{GW}$ is the one that is usually considered
in the literature, it would be of great interest to know, whether 
non-ultralocality of GW actions extends to the more general case.

Contrary to the weak non-ultralocality, which holds even in the presence 
of fermion doubling, non-ultralocality of actions can only hold for 
doubler--free actions. This is because (at least in free case) there
are infinitely many chirally nonsymmetric ultralocal GW actions with
doublers, e.g.
\begin{displaymath}
   D(p) \;=\; \sum_\mu \sin^2 p_\mu \, \identity + i\sin p_\mu \,\gmu
         \qquad
        \bigl(D^{-1}\bigr)_N \;=\; 
        { 1 \over 1+\sum_\mu \sin^2 p_\mu} \; \identity        
\end{displaymath}

Consequently, at the free level, the hypothesis of {\it ``strong
non-ultralocality''} can be formulated like this~\cite{Hor99A} 

\smallskip
\begin{itemize}
  \item[] {\it HYPOTHESIS:
    There is no $D(p)\in A$ such that the following three
    requirements are satisfied simultaneously:   
    \begin{description}
      \item[$(\alpha)$] $D(p)$ involves finite number of Fourier terms. 
      \item[$(\beta)$] $\bigl(D^{-1}(p)\bigr)_N$ is analytic.
      \item[$(\gamma)$] $\bigl(D^{-1}(p)\bigr)_C$ has no poles except if
                        $\,p_\mu=0\!\pmod{2\pi},\, \forall\mu$. 
    \end{description}}
\end{itemize}
Conditions $(\alpha-\gamma)$ represent ultralocality, GWL symmetry,
and the absence of doublers. Below I will describe an algebraic problem
which, I believe, holds the key to this issue. My reasoning will 
necessarily be terse, but the resulting problem will be stated 
clearly.

If non-ultralocality indeed holds, it will most likely result from
the clash of the two analyticity properties $(\beta), (\gamma)$. I will
consider the two-dimensional restrictions of the lattice Dirac operators
in higher (even) dimensions, because they are already capable of 
capturing the required analytic structure. As a result of hypercubic 
symmetry, the restrictions have the form (the term proportional
to $\gamma_1\gamma_2$ is ignored for simplicity)
\begin{displaymath}
   D(p) \;=\; A(p)\, \identity + i B_\mu (p)\, \gmu \qquad\quad
   D^{-1} \;=\; { A\, \identity - i B_\mu\, \gmu \over
                         A^2 + B_\mu B_\mu }                          
\end{displaymath}        
where $p=(p_1,p_2)$, $\mu=1,2$ and the functions $A(p), B_\mu(p)$
have the appropriate symmetry properties.

The crucial difference between ultralocal and non-ultralocal actions is
that in the former case we only have finite number of coefficients to 
adjust so that $(\beta),(\gamma)$ are satisfied, while in the latter case
there are infinitely many. This is more explicit if one makes the change 
of variables, such as  
$\; x = \sin{ p_1\over 2}, \; y = \sin{ p_2\over 2}$, 
which does not change the analytic structure of the relevant functions.
Then we are essentially dealing with polynomials.

It is easy to see that requirement $(\beta)$ is particularly restrictive
because it implies that the symmetric rational function 
$R(x^2,y^2) \equiv A/(A^2+B_\mu B_\mu)$ is analytic on the domain
$[-1,1]\times [-1,1]$, while the polynomial $A^2+B_\mu B_\mu$ vanishes
at the origin. This is only possible if the numerator and denominator
have a common polynomial factor which can be canceled so that the
denominator does not vanish anymore. From the structure of $R(x^2,y^2)$
it follows that $A(x^2,y^2)$ and $B(x^2,y^2)\equiv B_\mu B_\mu$ 
must each have this polynomial factor. It turns out that apart from 
the necessary zero at the origin, such common factors $F(x,y)$ tend to 
possess another zero in the domain $[-1,1]\times [-1,1]$, which then
makes the inclusion of requirement $(\gamma)$ impossible. Consequently,
it would be inherently useful to prove or disprove the following hypothesis:

\begin{itemize}
  \item[] {\it HYPOTHESIS: Let $G$ be the polynomial in $x^2,y^2$ 
  with complex coefficients, such that $G(0,0)=1$. There is no 
  $G$ such that the polynomial
  \begin{displaymath}
    B(x^2,y^2) \;=\; 4x^2(1-x^2)\; G^2(x^2,y^2) + 
                     4y^2(1-y^2)\; G^2(y^2,x^2)
  \end{displaymath}
  can be factorized as $B(x^2,y^2) = P(x,y)F(x,y)$, where 	
  $P(0,0)\ne 0$, $F(0,0)=0$, and $F(x,y)\ne 0$ elsewhere
  on the domain $[-1,1]\times [-1,1]$.
  }
\end{itemize}
The above form of $B(x^2,y^2)$ is dictated by hypercubic
symmetries. I stress that if this hypothesis is true, then it
implies ``strong non-ultralocality'' of GW actions. On the other 
hand, the possible examples of ultralocal GW actions can only be 
built out of counterexamples to this algebraic statement.

\section{Simple GW Actions?}

Assuming that GW actions indeed can not be simple in position space
(non-ultralocality), one naturally asks what kind of other practically 
useful properties they {\it can have}. I propose to examine the 
possibility that GW actions can be simple in eigenspace.

If the complete left-right eigenset 
$\{\;|\,\phi^i_L(U)>,\, |\,\phi^i_R(U)>, \lambda_i(U) \,\}$ exists
for $D(U)\in A$, then we can represent the operator as
\begin{displaymath}
  D \;\,=\;\, \sum_i \; |\, \phi_R^i > \;\lambda_i\; < \phi_L^i\,| 
\end{displaymath}
This representation is useful even in case of Wilson and staggered 
fermions, since the effects of light quarks are quickly accounted for 
by including only the lightest eigenmodes in the sum. The underlying 
idea is appealing for both generation of dynamical 
configurations~\cite{Dun99A,Hor99B}, and for propagator 
technology~\cite{Neg98A}: once the approximate eigenmode
representation is computed, the resulting quark propagators can be
tied together in any way desired. The group at the University of 
Virginia is currently actively pursuing this approach (see also the 
talk by T.~Lippert).

The eigenspace representation of the operator $D(U)$ is ``simple'', 
if the corresponding eigenbasis can be calculated efficiently. Even if
$D(U)$ is not ultralocal, there still may be a {\it commuting 
ultralocal} operator $Q(U)$ with the same eigenbasis. In this case
the eigenspace representation of $D$ is as simple as the eigenspace 
representation of $Q$. Consequently, it would be very interesting 
to know whether there are local, doubler-free elements of $A^{GW}$,
for which such an ultralocal operator $Q$ exists.

Obvious candidates for GW actions of the above type would be the
functions of the ultralocal operator $Q$, i.e. $D=F(Q)$. 
It is an open question whether such GW actions exist.
Another simple possibility is to consider the functions $F(Q,Q^+)$,
where $Q$ is the ultralocal normal operator $[Q,Q^+]=0$. The task
of finding such actions simplifies a lot if one only considers
the operators $Q=D_0$ representing valid, doubler-free lattice 
Dirac operator with $\gfive$-hermiticity ($D_0^+ = \gfive D_0 \gfive$). 
This is because to such $D_0\in A^{ND}$ one can directly assign 
a doubler-free element $D\in A^{GW}$, in a way analogous to the
Neuberger construction~\cite{Neu98B}, i.e.
\begin{displaymath}
    D = m_0 \Bigl[ 1 + \bigl(D_0-m_0) 
    {1 \over \sqrt{(D_0-m_0)^+ (D_0-m_0)}} \Bigr]
\end{displaymath}
with appropriate choice of $m_0$.
One is thus lead to consider the following:
\begin{itemize}
  \item[] {\it PROBLEM: Are there any ultralocal elements 
   $D_0\in A^{ND}$ with $\gfive$-hermiticity that are normal?}
\end{itemize}
This is a beautiful problem with trivial solutions at the free level 
(e.g. Wilson-Dirac operator), but none are known in arbitrary gauge 
background.

To get a flavour of what is involved here, it is useful to unmask
the spinorial structure of the problem. For example, in two dimensions
any operator $D\in A$ has the form 
$ D = A \identity + i B_\mu \gmu + C \gfive$ where $A,B_\mu,C$ are gauge 
invariant matrices with position and gauge indices only. 
$\gfive$-hermiticity implies that $A,B_\mu,C$ are hermitian,
and in this case normality demands
\begin{equation}
   \{ B_\mu, C \} \;=\; i \epsilon_{\mu\nu} [B_\nu, A]
  \label{eq:5} 
\end{equation}
The challenge is to find out whether, having only finite number of gauge
paths at our disposal, we can arrange for the above identities to hold. 
This is quite nontrivial and definite properties of $A,B_\mu,C$ 
under hypercubic transformations represent an important constraint here. 

I would also like to point out that in the above language, it is easy to 
understand how $\gfive$-hermiticity combined with normality simplifies 
the algebraic structure imposed by GW symmetry. 
Indeed, if one identifies $J_1 \equiv A-1, J_2\equiv B_1, J_3\equiv B_2$, 
and $C^2 = 1 - J_\mu J_\mu$, then the canonical GW relation 
$\{D,\gfive\} = D\gfive D$ in two dimensions translates into
\begin{equation}
    \{ J_\mu, C \} \;=\; i \epsilon_{\mu\nu\rho} [ J_\nu, J_\rho ]
    \label{eq:10}
\end{equation}
Relations (\ref{eq:5}) form a subset of the above identities that are 
automatically satisfied if $\gfive$-hermiticity and normality are 
demanded. The algebraic structure (\ref{eq:10}) implied by GW relation 
is  perhaps interesting by itself and deserves further study.

Finally, I would like to introduce the lattice Dirac operator which might
be of practical relevance in the context of using the eigenspace 
techniques in lattice QCD. 
Let $\{\,|\,\phi^i_L(U)>,\, |\,\phi^i_R(U)>, \lambda_i(U)\,\}$ be the 
eigenset of the Wilson-Dirac operator $D_W(U)$, and let $m_0\in (0,2)$.
Consider the operator
\begin{equation}
  D \;=\; \sum_i \; |\, \phi_R^i > \;
    m_0 \Bigl[ 1 + {{\lambda_i-m_0} \over 
    \sqrt{(\lambda_i-m_0)^* (\lambda_i-m_0)}} \Bigr] 
    \; < \phi_L^i\,|
    \label{eq:15}
\end{equation}   
This is a well defined operator for arbitrary gauge background in which 
the left-right eigenbasis of $D_W$ exists. In trivial gauge background 
($U\rightarrow 1$) it coincides with the Neuberger operator, and the
spectrum allways lies on a circle with radius $m_0$. While the locality 
properties of $D$ are questionable, the fact that it is perfectly local 
in the free limit suggests that non-local parts, which might be present, 
will be  arbitrarily small on sufficiently smooth backgrounds. Even 
though this can certainly cause practical concerns at intermediate 
couplings, it would seem unlikely that there is a problem of principle 
as the continuum limit is approached. 

Operator (\ref{eq:15}) should have improved chiral properties relative 
to the Wilson-Dirac operator, while its computational demands in the 
eigenspace approach are approximately the same. The degree to 
which the chirally non-symmetric part of the propagator is local in 
nontrivial backgrounds (it is proportional to a delta function in
free limit) is an open question. At the same time, however, the fact 
that the spectrum is forced on a circle suggests that the additive 
mass renormalization will be small (if any). These issues are currently
under investigation.

\section*{Acknowledgment} 

I thank Hank Thacker and Ziad Maassarani for many pleasant discussions 
on the topics presented here. I would also like to express my gratitude 
to all the organizers of this Workshop for excellent hospitality.

\end{document}